\documentclass[prd,preprint,superscriptaddress,showpacs,byrevtex]{revtex4}
\usepackage{bm}
\def\fsl#1{\setbox0=\hbox{$#1$}           % set a box for #1
   \dimen0=\wd0                                 % and get its size
   \setbox1=\hbox{/} \dimen1=\wd1               % get size of /
   \ifdim\dimen0>\dimen1                        % #1 is bigger
      \rlap{\hbox to \dimen0{\hfil/\hfil}}      % so center / in box
      #1                                        % and print #1
   \else                                        % / is bigger
      \rlap{\hbox to \dimen1{\hfil$#1$\hfil}}   % so center #1
      /                                         % and print /
   \fi}                                         %
\newcommand{\be}{\begin{equation}}
\newcommand{\ee}{\end{equation}}
\newcommand{\bea}{\begin{eqnarray}}
\newcommand{\eea}{\end{eqnarray}}
\newcommand{\beq}{\begin{equation}}
\newcommand{\eeq}{\end{equation}}
\newcommand{\beqs}{\begin{eqnarray}}
\newcommand{\eeqs}{\end{eqnarray}}

\newcommand{\pslash}{p\hspace{-0.067in}\slash}

\newcommand{\Aslash}{A\hspace{-0.067in}\slash}
\begin{document}
\title{ Altarelli-Parisi Equation in Non-Equilibrium QCD }
\author{Gouranga C Nayak } \email{nayak@physics.arizona.edu}
\affiliation{Department of Physics, University of Arizona, Tucson, AZ 85721, USA }
\begin{abstract}
The $Q^2$ evolution of fragmentation function in non-equilibrium
QCD by using DGLAP evolution equation may be necessary to study hadron formation
from quark-gluon plasma at RHIC and LHC.
In this paper we study splitting functions in non-equilibrium QCD by
using Schwinger-Keldysh closed-time path integral formalism.
For quarks and gluons with arbitrary non-equilibrium distribution functions $f_q({\vec p})$
and $f_g({\vec p})$, we derive expressions for quark and gluon splitting functions
in non-equilibrium QCD at leading order in $\alpha_s$. We make a comparison of these
splitting functions with that obtained by Altarelli and Parisi in vacuum.
\end{abstract}
\maketitle
%
%\newpage
%
\pagestyle{plain}
\pagenumbering{arabic}
\section{Introduction}
RHIC and LHC heavy-ion colliders are the best facilities to study quark-gluon
plasma in the laboratory. Since two nuclei travel almost at speed of light,
the QCD matter formed at RHIC and LHC may be in non-equilibrium. In order
to make meaningful comparison of the theory with the experimental data on hadron
production, it may be necessary to study nonequilibrium-nonperturbative QCD at
RHIC and LHC. This, however, is a difficult problem.

Non-equilibrium quantum field theory can be studied by using Schwinger-Keldysh closed-time
path (CTP) formalism \cite{schw,keldysh}. However, implementing CTP in non-equilibrium at
RHIC and LHC is a very difficult problem, especially due to the presence of gluons in
non-equilibrium and hadronization etc. Recently, one-loop resumed gluon propagator in
non-equilibrium in covariant gauge is derived in \cite{greiner,cooper}.

High $p_T$ hadron production at high energy $e^+e^-$, $ep$ and $pp$ colliders is studied
by using Collins-Soper fragmentation function \cite{collins,frag}.
For a high $p_T$ parton fragmenting to hadron, Collins-Soper derived an expression
for the fragmentation function based on field theory and factorization properties in QCD at
high energy \cite{george}. This fragmentation function is universal in the sense that, once
its value is determined from one experiment it explains the data at other experiments.

Recently we have derived parton to hadron fragmentation function in non-equilibrium QCD
by using Schwinger-Keldysh closed-time path integral formalism \cite{nayakfrag}.
This can be relevant at RHIC and LHC heavy-ion colliders to study hadron production
from quark-gluon plasma. We have considered a high $p_T$ parton in medium at
initial time $\tau_0$ with arbitrary non-equilibrium (non-isotropic) distribution function
$f(\vec{p})$ fragmenting to hadron. The special case $f(\vec{p})=\frac{1}{e^{\frac{p_0}{T}}\pm 1}$
corresponds to the finite temperature QCD in equilibrium.

We have found the following definition of the parton to hadron fragmentation function
in non-equilibrium QCD by using closed-time path integral formalism \cite{nayakfrag}.
For a quark ($q$) with arbitrary non-equilibrium distribution function $f_q(\vec{k})$ at
initial time, the quark to hadron fragmentation function is given by
\bea
&& D_{H/q}(z,P_T)
= \frac{1}{2z~[1+f_q({\vec k})]} \int dx^- \frac{d^{d-2}x_T}{(2\pi)^{d-1}}  e^{i{k}^+ x^- + i {P}_T \cdot x_T/z} \nonumber \\
 &&~\frac{1}{2}{\rm tr_{Dirac}}~\frac{1}{3}{\rm tr_{color}}[\gamma^+<in| \psi(x^-,x_T) ~\Phi[x^-,x_T]~a^\dagger_H(P^+,0_T)  a_H(P^+,0_T) ~\Phi[0]~{\bar \psi}(0)  |in>]
\label{qnf}
\eea
where $z$ (=$\frac{P^+}{k^+}$) is the longitudinal momentum fraction of the hadron with respect to the
parton and $P_T$ is the transverse momentum of the hadron.
For a gluon ($g$) with arbitrary non-equilibrium distribution function $f_g(\vec{k})$ at
initial time, the gluon to hadron fragmentation function is given by
\bea
&& D_{H/g}(z,P_T)
= \frac{1}{2zk^+~[1+f_g({\vec k})]} \int dx^- \frac{d^{d-2}x_T}{(2\pi)^{d-1}}  e^{i{k}^+ x^- + i {P}_T \cdot x_T/z} \nonumber \\
&&~\frac{1}{8} \sum_{a=1}^8 [<in| F^{+\mu}_a(x^-,x_T) ~\Phi[x^-,x_T]~a^\dagger_H(P^+,0_T)  a_H(P^+,0_T)~\Phi[0]~  F^+_{\mu a}(0) |in>].
\label{gnf}
\eea
In the above equations $|in>$ is the initial state of the non-equilibrium quark (gluon)
medium. The path ordered exponential
\bea
\Phi[x^\mu ]={\cal P}~ {\rm exp}[ig\int_{-\infty}^0 d\lambda~ n \cdot A^a(x^\mu +n^\mu \lambda )~T^a]
\label{wilf}
\eea
is the Wilson line \cite{collins,george,tucci}.

Eqs. (\ref{qnf}) and (\ref{gnf}) can be compared with the following definition of Collins-Soper
fragmentation function in vacuum \cite{collins}:
\bea
&& D_{H/q}(z,P_T)
= \frac{1}{2z} \int dx^- \frac{d^{d-2}x_T}{(2\pi)^{d-1}}  e^{i{k}^+ x^- + i {P}_T \cdot x_T/z} \nonumber \\
 &&~\frac{1}{2}{\rm tr_{Dirac}}~\frac{1}{3}{\rm tr_{color}}[\gamma^+<0| \psi(x^-,x_T) ~\Phi[x^-,x_T]~a^\dagger_H(P^+,0_T)  a_H(P^+,0_T) ~\Phi[0]~{\bar \psi}(0)  |0>]
\label{qvf}
\eea
and
\bea
&& D_{H/g}(z,P_T)
= -\frac{1}{2zk^+} \int dx^- \frac{d^{d-2}x_T}{(2\pi)^{d-1}}  e^{i{k}^+ x^- + i {P}_T \cdot x_T/z} \nonumber \\
&&~\frac{1}{8} \sum_{a=1}^8 [<0| F^{+\mu}_a(x^-,x_T) ~\Phi[x^-,x_T]~a^\dagger_H(P^+,0_T)  a_H(P^+,0_T)~\Phi[0]~  F^+_{\mu a}(0) |0>].
\label{gvf}
\eea

Since the fragmentation function is a non-perturbative quantity, we do not have
theoretical tools in QCD to calculate it yet. The normal procedure at high energy
$pp$, $ep$ and $e^+e^-$ colliders is to extract it at some initial momentum scale
$\mu_0$ and then evolve it to another scale $\mu$ by using the DGLAP evolution
equation \cite{gl,ap,d}:
\bea
\mu \frac{\partial }{\partial \mu}  D_{i \rightarrow j/\psi} (z) = \sum_j \int_z^1
\frac{dy}{y} P_{ij}(\frac{z}{y},\mu) D_{j \rightarrow j/\psi} (y).
\label{ap}
\eea
In the above equation $P_{ij}(z)$ is the splitting function of a parton $j$ into a
parton $i$ which is related to the probability of a parton $j$ emitting a parton
$i$ with longitudinal momentum fraction $z$. The quark and gluon splitting functions
$P_{ij}(z)$ in vacuum is evaluated by Altarelli
and Parisi in \cite{ap} at the leading order in coupling constant $\alpha_s$.

In order to apply this procedure at high energy heavy-ion colliders at RHIC and
LHC one needs to prove factorization of fragmentation function in non-equilibrium QCD.
Recently we have proved factorization theorem in non-equilibrium QED in \cite{fqed}
and in non-equilibrium QCD in \cite{fqcd}.

In this paper we will evaluate the quark and gluon splitting functions in non-equilibrium
QCD at the leading order in coupling constant $\alpha_s$ by using closed-time path integral
formalism. We find that these splitting functions
depend on non-equilibrium distribution functions of quarks and gluons in the QCD medium.
The $Q^2$ evolution of the non-equilibrium fragmentation functions (eqs. (\ref{qnf}) and
(\ref{gnf})) can be studied from eq. (\ref{ap}) by using non-equilibrium splitting
functions.

We find the following expressions for the quark and gluon splitting functions in non-equilibrium QCD
at leading order in coupling constant $\alpha_s$
\bea
&& P_{gq}(z) = C_2(R) ~[1+f_q(k)]^2~[1+f_g(k_T,zk)]^2~[1+f_q(-k_T, (1-z)k)]^2~[\frac{1+(1-z)^2}{z}] \nonumber \\
&& P_{qq}(z) = C_2(R) ~[1+f_q(k)]^2~[1+f_g(-k_T,(1-z)k)]^2~[1+f_q(k_T, zk)]^2~[\frac{1+z^2}{1-z}] \nonumber \\
&& P_{gg}(z) = 2C_A ~[1+f_g(k)]^2~[1+f_g(k_T,zk)]^2~[1+f_g(-k_T, (1-z)k)]^2 \nonumber \\
&&~[\frac{1-z}{z} + \frac{z}{1-z}+z(1-z)].
\label{65mi}
\eea
where $k$ is the momentum of initial parton (which is assumed to be along longitudinal
direction), $k_T$ is the transverse momentum of the emitted parton and $z$ is the
longitudinal momentum fraction of the initial parton carried by the emitted parton.

Eq. (\ref{65mi}) can be compared with the following expressions for the splitting
functions in vacuum obtained by Altarelli and Parisi \cite{ap} at the leading order
in coupling constant $\alpha_s$:
\bea
&& P_{gq}(z) = C_2(R) ~\frac{1+(1-z)^2}{z} \nonumber \\
&& P_{qq}(z) = C_2(R) ~\frac{1+z^2}{1-z} \nonumber \\
&& P_{gg}(z) = 2C_A ~[\frac{1-z}{z} + \frac{z}{1-z}+z(1-z)].
\label{65m0}
\eea

We will present derivation of eq. (\ref{65mi}) in this paper.

The paper is organized as follows. In section II we briefly review the derivation of
quark and gluon splitting functions in vacuum. In section III we describe Schwinger-Keldysh
closed-time path integral formalism in non-equilibrium QCD relevant to our calculation. In
section IV we derive quark and gluon splitting functions $P_{ij}$ in non-equilibrium QCD
by using closed-time path integral formalism. Section V contains conclusions.

\section{ Quark and Gluon Splitting Functions in Vacuum }

In this section we briefly review the derivation of quark and gluon
splitting functions $P_{ij}$ in vacuum. We will present our calculation
in the $S-$matrix approach. Hence, our derivation is slightly different
from  \cite{ap}.

Consider a quark with momentum $p_A$ emitting a gluon with momentum $p_B$
in the process $q (p_A) \rightarrow g (p_B) + q (p_C)$.
The $S$-matrix element for this process is given by
\bea
S^{(1)}=ig \int d^4x N[{\bar \psi}(x) \Aslash^a(x)T^a \psi(x)]
\label{s1}
\eea
where (the normalization is from \cite{mandl})
\bea
&& ~\psi(x) = \psi^+(x) + \psi^-(x) = \sum_{\rm spin} \sum_p \sqrt{\frac{m}{VE_p}}
 [a_q(p) u(p) e^{-ip\cdot x} + a^\dagger_{\bar q}(p)
v(p) e^{ip\cdot x} ]\nonumber \\
&& ~{\bar \psi }(x) = {\bar \psi}^+(x) + {\bar \psi}^-(x)
= \sum_{\rm spin} \sum_p \sqrt{\frac{m}{VE_p}}
[a_{\bar q}(p) {\bar v}(p) e^{-ip\cdot x} + a^\dagger_q(p) {\bar u}(p) e^{ip\cdot x}] \nonumber \\
&& ~A^\mu(x) = A^{\mu +}(x) + A^{\mu -}(x)
= \sum_{\rm spin} \sum_p \sqrt{\frac{1}{2VE_p}}
[a_g(p) \epsilon^\mu(p) e^{-ip\cdot x} + a^\dagger_g(p) \epsilon^\mu(p) e^{ip\cdot x}].
\label{wf}
\eea
In the above equation $a_q(p)$, $a_{\bar q}(p)$ and $a_g(p)$ are annihilation operators for
quark, antiquark and gluons respectively. In eq. (\ref{wf}) the suppression of color indices
are understood. The initial and final states are
\bea
|i> = |q(p_A)>=a^\dagger_q(p_A) |0>,~~~~~~~~~~~~~~~~~|f> = |q(p_C),~g(p_B)>=a^\dagger_q(p_C) a^\dagger_g(p_B) |0>,
\label{inout0}
\eea
where $p_C=p_A-p_B$. Hence we find
\bea
|<f|S^{(1)}|i>|^2 =
[\frac{V}{(E_C+E_B-E_A)}]^2~\frac{m}{VE_{C}}\frac{m}{VE_{{A}}}\frac{1}{2VE_{{B}}} ~\sum_{\rm spin} |M|^2.
\label{fs1i6}
\eea
where
\bea
M=ig{\bar u}(p_C) \gamma_\mu u(p_{A})\epsilon^{a \mu}(p_{B}) T^a.
\label{matr}
\eea

For massless quarks we find
\bea
&&W_{gq} = |<f|S^{(1)}|i>|^2 \frac{V d^3p_C}{(2\pi)^3}\nonumber \\
&&= C_2(R)g^2  \frac{d^3p_C}{(2\pi)^3} \frac{E_B}{2E_{A}E_{C}(2E_B)^2} \frac{1}{(E_C+E_B -E_A)^2}
~{\rm Tr}[{\pslash }_C \gamma^i {\pslash }_A \gamma^j ]~ (\delta^{ij}-\frac{p_B^i p_B^j}{{\vec{p}_B^2}})
\label{prob3}
\eea
which gives the quark to gluon splitting function 
\bea
P_{gq}(z) = C_2(R) \frac{1+(1-z)^2}{z}.
\label{59a}
\eea
Similarly we find the quark to quark splitting function 
\bea
P_{qq}(z) = C_2(R)~\frac{1+z^2}{1-z}
\label{60}
\eea
and the gluon to gluon splitting function 
\bea
P_{gg}(z) = 2C_A ~[\frac{1-z}{z} + \frac{z}{1-z}+z(1-z)].
\label{65}
\eea

\section{ Non-equilibrium QCD Using Closed-Time Path Formalism }

Unlike $pp$ collisions, the ground state at RHIC and LHC heavy-ion collisions
(due to the presence of a QCD medium at initial time $t=t_{in}$ (say $t_{in}$=0)
is not a vacuum state $|0>$ any more. We denote $|in>$ as the initial state of the non-equilibrium QCD
medium at $t_{in}$. The non-equilibrium distribution function $f(\vec{k})$ of a parton (quark or gluon),
corresponding to such initial state is given by
\bea
<a^\dagger ({\vec k})a({\vec k}')>=<in|a^\dagger ({\vec k})a({\vec k}')|in> = f(\vec{k})
\delta^{(3)}_{{\vec k}{\vec k}'}
\label{dist}
\eea
where we have assumed space translational invariance at initial time.

Finite temperature field theory formulation is a special case of this when
$f({\vec k}) =\frac{1}{e^{\frac{k_0}{T}} \pm 1}$.

\subsection{ Quarks in non-Equilibrium }

The non-equilibrium (massless) quark propagator at initial time $t=t_{in}$ is given by
(suppression of color indices are understood)
\bea
G(k)_{ij}=\displaystyle{\not}k \left ( \begin{array}{cc}
\frac{1}{k^2+i\epsilon}+2\pi \delta(k^2) f_q(\vec{k}) & -2\pi \delta(k^2)\theta(-k_0)+2\pi \delta(k^2) f_q(\vec{k}) \\
-2\pi \delta(k^2)\theta(k_0)+2\pi \delta(k^2) f_q(\vec{k}) & -\frac{1}{k^2-i\epsilon}+2\pi \delta(k^2) f_q(\vec{k})
\end{array} \right )
\eea
where where $i,j= +,-$ and $f_q(\vec{k})$ is the arbitrary non-equilibrium distribution function of quark.

\subsection{ Gluons in Non-Equilibrium }

We work in the frozen ghost formalism \cite{greiner,cooper} where the non-equilibrium
gluon propagator at initial time $t=t_{in}$ is given by (the suppression of color
indices are understood)
\bea
G^{\mu \nu}(k)_{ij} = -i[g^{\mu \nu} +  (\alpha -1) \frac{k^\mu k^\nu}{k^2}] ~G^{\rm vac}_{ij}(k) -iT^{\mu \nu}G^{\rm med}_{ij}(k)
\label{gpm}
\eea
where $i,j= +,-$. The transverse tensor is given by
\bea
T^{\mu \nu} (k)=g^{\mu \nu} -\frac{(k \cdot u)(u^\mu k^\nu + u^\nu k^\mu)-k^\mu k^\nu -k^2u^\mu u^\nu}{(k \cdot u)^2 -k^2}
\label{tmn}
\eea
with the flow velocity of the medium $u^\mu$. $G^{\mu \nu}_{ij}(k)$ are the usual vacuum propagators of the gluon
\bea
G^{\rm vac}_{ij}(k)=
\left ( \begin{array}{cc}
\frac{1}{k^2+i\epsilon} & -2\pi \delta(k^2)\theta(-k_0) \\
-2\pi \delta(k^2)\theta(k_0) & -\frac{1}{k^2-i\epsilon}
\end{array} \right )
\eea
and the medium part of the propagators are given by
\bea
G^{\rm med}_{ij}(k)= 2\pi \delta(k^2) f_g(\vec{k})
\left ( \begin{array}{cc}
1 & 1 \\
1 & 1
\end{array} \right ).
\eea

\subsection{ Ratio of Characteristic Relaxation Time of the Non-Equilibrium State to the QCD Evolution Time }

The typical relaxation time in the non-equilibrium QCD plasma can be written as \cite{qgp1,r1,r2}
\bea
\tau_c = \frac{1}{n {\hat \sigma}_{\rm tr} }
\label{1}
\eea
where
\bea
n=\int d^3k f({\vec k})
\label{2}
\eea
is the parton number density in terms of the non-equilibrium parton distribution function
$f({\vec k})$ and ${\hat \sigma}_{\rm tr}$ is the typical transport cross section of the partonic collisions
in the non-equilibrium QCD plasma which depends on the non-equilibrium parton distribution function
$f({\vec k})$.

Consider for example, the $gg \rightarrow gg$ scattering. The leading order partonic differential cross section
in vacuum is given by
\bea
\frac{d{\hat \sigma}}{d{\hat t}}=\frac{9 \pi \alpha_s^2}{2{\hat s}^2}[3-\frac{{\hat u}{\hat t}}{{\hat s}^2}-\frac{{\hat s}{\hat t}}{{\hat u}^2}-\frac{{\hat s}{\hat u}}{{\hat t}^2}]
\label{3}
\eea
which in the infrared limit ${\hat t} \rightarrow 0$ diverges
\bea
\frac{d{\hat \sigma}}{d{\hat t}}=\frac{9 \pi \alpha_s^2}{2{\hat t}^2}.
\label{4}
\eea
However, in the medium, the medium modified resumed gluon propagator removes this infrared divergence
and the typical finite differential cross section becomes \cite{nayak1}
\bea
\frac{d{\hat \sigma}}{d{\hat t}}=\frac{9 \pi \alpha_s^2}{8}[\frac{1}{(\Pi_L-{\hat t})({\bar \Pi}_L-{\hat t})}+\frac{1}{(\Pi_T-{\hat t})({\bar \Pi}_L-{\hat t})}+\frac{1}{(\Pi_L-{\hat t})({\bar \Pi}_T-{\hat t})}+\frac{1}{(\Pi_T-{\hat t})({\bar \Pi}_T-{\hat t})}]\nonumber \\
\label{5}
\eea
where $\Pi_L$ and $\Pi_T$ are the longitudinal and transverse component of the gluon self energy which depend
on the non-equilibrium distribution function $f({\vec k})$. One can see that even at the one-loop level
of the self energy the magnetic screening mass is non-zero \cite{greiner,cooper,kao1,kao2} as long
as the non-equilibrium distribution function $f({\vec k})$ is non-isotropic {\it i.e.}, it depends
on the direction of ${\vec k}$ of the parton, which is the case at early stage of the heavy-ion collisions
at RHIC and LHC. The expressions of the medium modified resumed gluon propagator at the one-loop level
of self energy in non-equilibrium in covariant gauge is recently derived in \cite{greiner,cooper}.

The transport cross section \cite{qgp1,nayak1}
\bea
{\hat \sigma}_{\rm tr} = \int_{-\frac{{\hat s}}{2}}^0  d{\hat t} \frac{d{\hat \sigma}}{d{\hat t}} {\rm sin}^2\theta_{\rm cm}= \int_{-\frac{{\hat s}}{2}}^0   d{\hat t} \frac{d{\hat \sigma}}{d{\hat t}} \frac{4 {\hat u}{\hat t}}{{\hat s}^2}.
\label{6}
\eea
for this process can be obtained by using eq. (\ref{5}). Since gluons are dominate part of the total parton
production at the early stage of the heavy-ion collisions at RHIC and LHC one can get an estimate of the 
relaxation time of the non-equilibrium state from eqs. (\ref{1}), (\ref{2}), (\ref{5}) and (\ref{6}).
For example, the typical value of the maximum relaxation time in non-equilibrium state found in \cite{qgp1} 
is $\sim $ 1.5 fm at RHIC and LHC heavy-ion colliders.

The typical QCD evolution time associated with the DGLAP evolution equation of the fragmentation
function is given by
\bea
t ={\rm ln}Q
\label{7}
\eea
where
\bea
Q=\mu
\label{8}
\eea
is the energy scale determined by the hard process probing fragmentation function \cite{QCDtime}.

\section{ Quark and Gluon Splitting Functions in Non-Equilibrium QCD }

In this section we evaluate quark and gluon splitting functions $P_{ij}$
in non-equilibrium QCD. Similar to the vacuum case in \cite{ap} (see eq. (\ref{inout0}))
we define the state $|i>$ and $|f>$ in non-equilibrium QCD as follows
\bea
|i> = |q(p_A)>= a^\dagger_q(p_A) |in>,~~~~~~~~~~~~~~~|f>= |q(p_C),~g(p_B)> = a^\dagger_q(p_C) a^\dagger_g(p_B) |in>
\label{inout0m}
\eea
where $|in>$ is the initial state of the non-equilibrium QCD medium. It has to be remembered that for evaluating the
Feynman diagrams and $S-$matrix we work in the interaction picture where the fields $\psi(x)$ and $A_\mu (x)$ obey
the free field equations in terms of creation and annihilation operators as given by eq. (\ref{wf}). From eqs.
(\ref{inout0m}) and (\ref{s1}) we find
\bea
<f|S^{(1)}|i> =ig \int d^4x <in|a_q(p_C) a_g(p_B) N[{\bar \psi}(x) \Aslash^{a }(x)T^a \psi(x)] a^\dagger_q(p_A) |in>
\label{fs1im}
\eea
which gives by using eq. (\ref{wf})
\bea
&& <f|S^{(1)}|i> =ig \int d^4x \sum_{\rm spin} \sum_{p,p',p''}\sqrt{\frac{m}{VE_p}}\sqrt{\frac{m}{VE_{p''}}}\sqrt{\frac{1}{2VE_{p'}}} <in|a_q(p_C) a_g(p_B) \nonumber \\
&&~[a^\dagger_q(p) {\bar u}(p) e^{ip\cdot x}] [\gamma_\mu a^\dagger_g(p') \epsilon^{a \mu}(p') e^{ip' \cdot x}T^a][ a_q(p'') u(p'') e^{-ip''\cdot x}] a^\dagger_q(p_A) |in>.
\label{fs1i1m}
\eea
Performing $x-$integration we find
\bea
&& <f|S^{(1)}|i> =ig \sum_{\rm spin} \sum_{p,p',p''}
\frac{V}{(E+E'-E'')}~\sqrt{\frac{m}{VE_p}}\sqrt{\frac{m}{VE_{p''}}}\sqrt{\frac{1}{2VE_{p'}}} <in|a_q(p_C) a_g(p_B) \nonumber \\
&&~[a^\dagger_q(p) {\bar u}(p) ] [\gamma_\mu a^\dagger_g(p') \epsilon^{a \mu}(p') T^a]
[ a_q(p'') u(p'')] a^\dagger_q(p_A) |in>.
\label{fs1i2m}
\eea
In the interaction picture the commutation relations are same as that for free field operators
\bea
[a(p), a^\dagger(p')] = \delta_{p p'}, ~~~~~~~~~~~~~~~~[a(p), a(p')] = [a^\dagger(p), a^\dagger(p')] =0.
\label{commm}
\eea
which gives
\bea
&& <f|S^{(1)}|i> =ig \sum_{\rm spin} \sum_{p,p',p''} \frac{V}{(E+E'-E'')}~\sqrt{\frac{m}{VE_p}}\sqrt{\frac{m}{VE_{p''}}}\sqrt{\frac{1}{2VE_{p'}}} <in| \nonumber \\
&&~[a^\dagger_q(p) a_q(p_C) +\delta_{p p_C}] {\bar u}(p) \gamma_\mu [a^\dagger_g(p') a_g(p_B)+\delta_{p' p_B}] \epsilon^{a \mu}(p') T^a [ a^\dagger_q(p_A)a_q(p'') +\delta_{p'' p_A}] u(p'')|in>. \nonumber \\
\label{fs1i3m}
\eea
For our purpose of evaluating Feynman diagrams in momentum space we use eq. (\ref{dist})
\bea
<in|a^\dagger_p a_{p'} |in> = f(\vec{p}) ~\delta^{(3)}_{\vec{p} \vec{p}'}
\label{iinn}
\eea
where we have assumed the space-translational invariance at initial time $t=t_{in}=0$.
Using eq. (\ref{iinn}) and summing over $p$, $p'$ and $p''$ we find from eq. (\ref{fs1i3m})
\bea
&& |<f|S^{(1)}|i>|^2 = [\frac{V}{(E_C+E_B-E_A)}]^2~\frac{m}{VE_{C}}\frac{m}{VE_{A}}\frac{1}{2VE_{{B}}} \nonumber \\
&&~\times~ [1+f_q({\vec{p}}_C)]^2~\times~[1+f_g(\vec{p}_B)]^2~\times~[1+f_q(\vec{p}_A)]^2~\sum_{\rm spin} |M|^2
\label{fs1i5m}
\eea
where
\bea
M=ig{\bar u}(p_C) \gamma_\mu u(p_{A})\epsilon^{a \mu}(p_{B}) T^a.
\label{matrm}
\eea

From now onwards we can follow exactly the same steps as in the vacuum case
(see section II, the derivations after eq. (\ref{fs1i6})) to find the probability
\bea
 W= C_2(R)\frac{\alpha_s}{ 2\pi}
~[1+f_q({\vec{p}}_C)]^2~\times~[1+f_g(\vec{p}_B)]^2~\times~[1+f_q(\vec{p}_A)]^2~ \frac{1+(1-z)^2}{z}~dz~d({\rm ln}p_T^2).\nonumber \\
\label{prob4m}
\eea

\subsection{ Quark to Gluon Splitting Function in Non-Equilibrium QCD }

From eq. (\ref{prob4m}) we find the quark to gluon splitting function in non-equilibrium
QCD at leading order in $\alpha_s$:
\bea
P_{gq}(z) = C_2(R) ~[1+f_q(p)]^2~[1+f_g(p_T,zp)]^2~[1+f_q(-p_T, (1-z)p)]^2~\frac{1+(1-z)^2}{z}
\label{59m}
\eea
which reproduces eq. (\ref{65mi}).

In the above equation $p$ is the momentum of initial quark (which is assumed to be along longitudinal
direction), $p_T$ is the transverse momentum of the gluon and $z$ is the longitudinal
momentum fraction of the initial quark carried by the gluon.

\subsection{ Quark to Quark Splitting Function in Non-Equilibrium QCD }

The quark to quark splitting function [in the process $q(p_A) \rightarrow q(p_B) + g(p_C)$]
can be obtained from the quark to gluon splitting function
[in the process $q(p_A) \rightarrow g(p_B) + q(p_C)$] with the replacement $z \rightarrow (1-z)$
\bea
P_{qq}(z) = P_{Gq}(1-z),~~~~~~~~~~~~~~~~~~~~z<1.
\eea
Hence we find from eq. (\ref{59m}) the quark to quark splitting function in non-equilibrium QCD
\bea
P_{qq}(z) = C_2(R) ~[1+f_q(p)]^2~[1+f_g(-p_T,(1-z)p)]^2~[1+f_q(p_T, zp)]^2~\frac{1+z^2}{1-z}
\label{60m}
\eea
which reproduces eq. (\ref{65mi}).

\subsection{ Gluon to Gluon Splitting Function in Non-Equilibrium QCD }

Similarly, using three gluon vertex and carrying out the similar algebra we
find gluon to gluon splitting function in non-equilibrium QCD
\bea
 P_{gg}(z) = 2C_A ~[1+f_g(p)]^2~[1+f_g(p_T,zp)]^2~[1+f_g(-p_T, (1-z)p)]^2~[\frac{1-z}{z} + \frac{z}{1-z}
+z(1-z)] \nonumber \\
\label{65m}
\eea
which reproduces eq. (\ref{65mi}).

The splitting functions in non-equilibrium QCD as given by eqs. (\ref{59m}), (\ref{60m}) and (\ref{65m}) can be
used to study DGLAP evolution equation of fragmentation function in non-equilibrium QCD \cite{nayakfrag,fqcd}
to study high $p_T$ hadron production from quark-gluon plasma at RHIC and LHC.

\section{Conclusions}

RHIC and LHC heavy-ion colliders are the best facilities to study quark-gluon
plasma in the laboratory. Since two nuclei travel almost at speed of light,
the QCD matter formed at RHIC and LHC may be in non-equilibrium.
Since the fragmentation function is a non-perturbative quantity, we do not have
theoretical tools in QCD to calculate it yet. The normal procedure at high energy
$pp$, $ep$ and $e^+e^-$ colliders is to extract it at some initial momentum scale
$\mu_0$ and then evolve it to another scale $\mu$ by using the DGLAP evolution
equation which involves splitting function $P_{ji}$ of a parton $j$ into a
parton $i$. The quark and gluon splitting functions in vacuum is evaluated by
Altarelli and Parisi in \cite{ap} at the leading order in coupling constant $\alpha_s$.
In order to apply this procedure at high energy heavy-ion colliders at RHIC and
LHC one needs to prove factorization of fragmentation function in non-equilibrium QCD.
Recently we have proved factorization theorem in non-equilibrium QED in \cite{fqed}
and in non-equilibrium QCD in \cite{fqcd}.

In this paper we have evaluated the quark and gluon splitting functions in non-equilibrium
QCD at the leading order in coupling constant $\alpha_s$ by using closed-time path integral
formalism.
For quarks and gluons with arbitrary non-equilibrium distribution functions $f_q({\vec p})$
and $f_g({\vec p})$, we have derived expressions for quark and gluon splitting functions
in non-equilibrium QCD. We have found that the quark and gluon splitting functions
depend on non-equilibrium distribution functions $f_q({\vec p})$ and $f_g({\vec p})$.
We have made a comparison of these
splitting functions with that obtained by Altarelli and Parisi in vacuum.

The splitting functions in non-equilibrium QCD can be
used to study DGLAP evolution equation of fragmentation function in non-equilibrium QCD \cite{nayakfrag,fqcd}
to study high $p_T$ hadron production from quark-gluon plasma \cite{qgp1,qgp2} at RHIC and LHC.

\acknowledgements
This work was supported in part by Department of Energy under
contracts DE-FG02-91ER40664, DE-FG02-04ER41319 and DE-FG02-04ER41298.


\begin{thebibliography}{99}
\bibitem{schw} J. Schwinger, J. Math. Phys. 2 (1961) 407.

\bibitem{keldysh} L. V. Keldysh, JETP 20 (1965) 1018.
\bibitem{greiner} C-W. Kao, G. C. Nayak and W. Greiner, Phys. Rev. D 66 (2002) 034017.

\bibitem{cooper} F. Cooper, C-W. Kao and G. C. Nayak, Phys. Rev. D 66 (2002) 114016;
hep-ph/0207370.

\bibitem{collins} J. C. Collins and D. E. Soper, Nucl. Phys, B 193 (1981) 381; Erratum-ibid.B213 (1983) 545;
Nucl. Phys. 194 (1982) 445.

\bibitem{frag} B. A. Kniehl, G. Kramer and B. Potter, Nucl.Phys.B582:514-536,2000;
S. Kretzer, Phys. Rev. D 62, 054001 (2000).

\bibitem{george} G. Sterman, Phys. Rev. D17, 2773 (1978); Phys. Rev. D17, 2789 (1978);
J. C. Collins, D. E. Soper and G. Sterman, Adv. Ser. Direct. High Energy Phys. 5, 1 (1988);
Nucl. Phys. B261, 104 (1985); Nucl. Phys. B 308 (1988) 833; G. T. Bodwin, S. J. Brodsky
and G. P. Lepage, Phys. Rev. Lett. 47, 1799 (1981); G. T. Bodwin, Phys. Rev. D31, 2616
(1985) [Erratum-ibid. D34, 3932 (1986)].

\bibitem{nayakfrag} G. C. Nayak, Eur. Phys. J.C59:891,2009.

\bibitem{tucci} R. Tucci, Phys. Rev. D32 (1985) 945.

\bibitem{gl} V. N. Gribov and L. N. Lipatov, Yad. Fiz. 15, 781 (1972) [Sov. J. Nucl. Phys.
15, 438 (1972)]; L. N. Lipatov, {\it ibid.} 20, 181 (1974) [{\it ibid.} 20, 94 (1975)].

\bibitem{ap} G. Altarelli and G. Parisi, Nucl. Phys. B126 (1977) 298.

\bibitem{d} Yu. L. Dokshitzer, Zh. Eksp. Teor. Fiz. 73, 1216 (1977) [Sov. Phys. JETP
46, 641 (1977)].

\bibitem{fqed} G. C. Nayak, Annals Phys.324:2579-2585,2009.

\bibitem{fqcd} G. C. Nayak, Annals Phys.325:682-690,2010.

\bibitem{mandl} F. Mandl and G. Shaw, {\it Quantum Field Theory}, John Wiley and Sons, 1984.

\bibitem{qgp1} G. C. Nayak {\it et al.}, Nucl. Phys. A687, 457 (2001).

\bibitem{r1} A. Hosoya and K. Kajantie, Nucl. Phys. B250, 666 (1985).

\bibitem{r2} P. Danielewicz and M. Gyulassy, Phys. Rev. D31, 53 (1985).

\bibitem{nayak1} J. Ruppert {\it et al.}, Phys. Lett. B 520, 233 (2001).

\bibitem{kao1} F. Cooper, C-W. Kao and G. C. Nayak, hep-ph/0207370.

\bibitem{kao2} M. C. Birse, C-W. Kao and G. C. Nayak, Phys. Lett. B570, 171 (2003).

\bibitem{QCDtime} K. Golec-Biernat, S. Jadach, W. Placzek and M. Skrzypek, Acta Physica Polonica B, Vol. 40, 1001 (2009).

\bibitem{qgp2} F. Cooper, E. Mottola and G. C. Nayak, Phys. Lett. B555, 181 (2003).

\end{thebibliography}
\end{document}